\documentclass[aps,pra,showpacs,twocolumn]{revtex4}

\usepackage{graphicx}
\begin{document}

\title{
All path-symmetric pure states achieve their maximal
phase sensitivity in conventional two-path interferometry}

\author{Holger F. Hofmann}
\email{hofmann@hiroshima-u.ac.jp}
\affiliation{
Graduate School of Advanced Sciences of Matter, Hiroshima University,
Kagamiyama 1-3-1, Higashi Hiroshima 739-8530, Japan}

\begin{abstract}
It is shown that the condition for achieving the quantum Cramer-Rao
bound of phase estimation in conventional two-path interferometers is
that the state is symmetric with regard to an (unphysical) exchange
of the two paths. Since path symmetry is conserved under phase shifts,
the maximal phase sensitivity can be achieved at arbitrary bias phases,
indicating that path symmetric states can achieve their quantum 
Cramer-Rao bound in Bayesian estimates of a completely unknown phase.
\end{abstract}


\maketitle

%%%--Introduction: qmetrology and two path interferometers, 

One of the most intriguing features of quantum physics is the
effect of quantization on the observation of classical interference 
effects. In conventional two-path interferometers such as the 
widely studied Mach-Zehnder interferometer, phase shifts are 
estimated from the intensity difference between the two output
ports. In quantum metrology, the phase shift is estimated from the
discrete particle statistics observed in the output. For $N$ 
uncorrelated particles, the randomness of the outcome results in a lower
bound of the phase sensitivity given by the shot noise limit (or standard
quantum limit) of $\delta \phi \geq 1/\sqrt{N}$.
However, quantum correlations between the particles can overcome this limit,
as was first demonstrated in the 1980s using squeezed light
\cite{Cav81,Xia87,Gra87}. Following this breakthrough, the theoretical
requirements for optimal phase estimates of general quantum states
were thoroughly studied \cite{Bra92,Bra94,San95,Bra96}, 
but the experimental technologies for
the detailed investigation of non-classical photon statistics were not
available at the time. 
Recently, there has been a renewed interest in quantum metrology due to
the first experimental realizations of maximally path entangled ``NOON''
states \cite{Wal04,Mit04} and the subsequent demonstrations of super phase
sensitivity using pair state inputs \cite{Nag07}, 
and due to the emerging possibilities of atom interferometry \cite{AtomInterf}. Consequently, new questions arise regarding the implications of the 
theoretical results for the recently developed experimental capabilities. 
In particular, it is a highly relevant practical question whether present 
photon counting experiments can provide an optimal phase estimation strategy
for a specific phase sensitive input state - that is, whether the
experiments can achieve the quantum Cramer-Rao bound of the state or not.

In general, the phase sensitivity of the experimental data will itself depend
on the bias phase, as confirmed by two recent studies of the experimental
phase sensitivities achieved by four photon pair states \cite{Oka08,Sun08}. 
These studies suggest that the Cramer-Rao bound can only be achieved around
specific phase shifts. However, the estimation strategies used only a 
subset of the possible output measurements. On the other hand, 
it has already been shown that the quantum Cramer-Rao bound can be achieved
at any phase bias in shot noise limited interferometry \cite{Pez07}.
As further work by the same group indicates, this result also applies to 
the states created by interference between squeezed light and coherent
light \cite{Pez08}.
 
In the following, it is shown that these important results can be generalized
to all pure states that are symmetric under a (non-physical) exchange of
the paths in the interferometer. This class of states covers most of
the states considered for non-classical phase measurements, including
maximally path entangled states \cite{Wal04,Mit04}, pair states 
\cite{Nag07,Oka08,Sun08}, the $N$-photon components
of coherent light and squeezed or down-converted light
\cite{Pez08,Hof07,Ono08}, and states generated
by photon subtraction \cite{Hof06}. As experimental methods improve, it should
therefore become possible to observe a phase independent uncertainty limited
phase sensitivity at the quantum Cramer-Rao bound for a wide range of 
non-classical input states.

%%%--Generator-estimator uncertainty
For the discussion of quantum properties of $N$-particle interferometry, 
it is convenient to express the quantum mechanics of two-paths interferometers
in terms of the spin-$N/2$ algebra of the Schwinger representation, 
\begin{eqnarray}
\label{eq:Jpar}
\hat{J}_1 = \frac{1}{2}(\hat{a}_1 ^\dagger \hat{a}_2
+\hat{a}_2 ^\dagger \hat{a}_1)
\nonumber \\
\hat{J}_2 = \frac{-i}{2}(\hat{a}_1 ^\dagger \hat{a}_2
-\hat{a}_2 ^\dagger \hat{a}_1)
\nonumber \\
\hat{J}_3 = \frac{1}{2}(\hat{a}_1 ^\dagger \hat{a}_1
-\hat{a}_2 ^\dagger \hat{a}_2),
\end{eqnarray}
where $\hat{a}_1$ and $\hat{a}_2$ are the annihilation operators of the 
paths inside the interferometer. A phase shift of $\Phi$ between the
arms of the interferometer can then be expressed by the unitary transformation
$\hat{U}(\Phi)=\exp(-i \Phi \hat{J}_3)$. Experimentally, the effects of
a small phase shift can be observed by measuring the average of an 
estimator observable $\hat{A}$. The differential change of this average 
is given by the expectation value of the commutation relation between
the generator $\hat{J}_3$ and the estimator $\hat{A}$,
\begin{equation}
\frac{\partial}{\partial \Phi} \langle \hat{A} \rangle
= -i \langle [\hat{A},\hat{J}_3] \rangle.
\end{equation}
Since the same commutation relation also determines the minimal product
of the uncertainties $\Delta J_3$ and $\Delta A$, the observable effects
of a small phase change are limited by the generator-estimator uncertainty,
\begin{equation}
\label{eq:geu}
\Delta J_3 \Delta A \geq \frac{1}{2}
\frac{\partial}{\partial \Phi} \langle \hat{A} \rangle.
\end{equation}
As pointed out in \cite{Bra94,Bra96}, this generalization of the Mandelstam-Tamm
uncertainty for energy and time directly defines the quantum Cramer-Rao
bound as
\begin{equation}
\label{eq:qcr}
\delta \phi^2 = \left( \frac{\Delta A}
{\frac{\partial}{\partial \phi} \langle \hat{A} \rangle} \right)^2
\geq \frac{1}{(2 \Delta J_3)^2}. 
\end{equation}
Thus the quantum Cramer-Rao bound of a pure state is given by its path 
uncertainty $\Delta J_3$. In principle, the bound can always be achieved if 
there are no restrictions of the possible measurements \cite{Bra94}.
However, conventional $N$-particle interferometry is limited
to the detection of the particle distribution in the two output ports. It is
therefore interesting to take a closer look at the properties of the 
specific estimators that minimize the uncertainty relation (\ref{eq:geu}).

%%%--Minimal uncertainty states
In Hilbert space, the generator-estimator uncertainty is based on the 
Cauchy-Schwarz inequality for the inner product of the vector 
$-i \hat{J}_3 \mid \psi \rangle$ representing the differential change 
of the quantum state and the vector $\hat{A}\mid \psi \rangle$
representing the uncertainty of the estimator. The quantum Cramer-Rao
bound is achieved if the estimator $\hat{A}$ satisfies the relation
\begin{equation}
\label{eq:state}
\lambda \hat{A} \mid \psi \rangle = -i \hat{J}_3 \mid \psi \rangle.
\end{equation}
It is easy to see that a large number of estimators fulfil this relation,
since it merely defines the matrix elements in one column and one line
of the $(N+1) \times (N+1)$ matrix describing $\hat{A}$ in any orthogonal
basis that includes $\mid \psi \rangle$. However, the situation changes
drastically when one requires that the eigenstates of the operator $\hat{A}$
must be given by the particle number states of the output ports. 
In terms of the spin-$N/2$ algebra defined in eq.(\ref{eq:Jpar}),
these are the $\hat{J}_1$-eigenstates $\{ \mid m_1 \rangle \}$. 
Eq.(\ref{eq:state}) then provides a unique definition of the eigenvalues
$A_m$ for each eigenstate $\mid m_1 \rangle$,
\begin{equation}
\label{eq:ev}
\lambda A_m = 
\frac{-i \langle m_1 \mid \hat{J}_3 \mid \psi \rangle
}{\langle m_1 \mid \psi \rangle}.
\end{equation}
Since the eigenvalues of the estimator observable $\hat{A}$ must be real,
the Cramer-Rao bound can only be achieved if the right hand side of
eq.(\ref{eq:ev}) is real as well. 

%%%--Path symmetry operation 
As can be understood from the cyclic properties of the spin algebra, the
matrix elements of  $\hat{J}_2$ in the $\hat{J}_3$-basis are all imaginary
(just like the matrix elements of $\hat{J}_3$ in the $\hat{J}_1$-basis).
Therefore, eq. (\ref{eq:ev}) results in real estimator values $A_m$ if
all quantum state components $\langle m_1 \mid \psi \rangle$ are real.
Specifically, the necessary and sufficient condition to obtain only real
$A_m$ for any state with non-zero components $\langle m_1 \mid \psi \rangle$
is that all components have the same phase factor $\chi_0$,
\begin{equation}
\label{eq:m1cond}
\langle m_1 \mid \psi \rangle = \langle m_1 \mid \psi \rangle^* 
e^{-2i\chi_0}.
\end{equation}
This condition can be interpreted as an invariance of the quantum state under
a symmetry operation. To visualize the physical meaning of this operation,
the complex conjugation of amplitudes in the $\hat{J}_1$-basis can be applied to
the operators $\hat{J}_i$. Since only the matrix elements of $\hat{J}_3$ are
real, the symmetry operation maps $\hat{J}_3$ to $-\hat{J}_3$ without changing
either $\hat{J}_1$ or $\hat{J}_2$. This means that the intensities in the
two paths of the interferometer are exchanged without changing the phase 
relation between the paths. Note that this unphysical exchange of paths is 
different from the one achieved by physically exchanging the modes in the two
arms, since such an exchange would flip either $\hat{J}_1$ or $\hat{J}_2$,
or a linear combination of the two specified by an appropriate reference phase.
On the other hand, the unphysical exchange of the paths does not depend on 
any reference phase between the arms of the interferometer. Hence, phase shifts
do not change path symmetry and a path symmetric quantum state will be path
symmetric at any bias phase $\phi$. 

The conservation of path symmetry under phase shifts can also be shown 
by transforming the symmetry condition of eq.(\ref{eq:m1cond}) into the 
$\hat{J}_3$-basis. The result reads
\begin{equation}
\label{eq:m3cond}
\langle m_3 \mid \psi \rangle = \langle - m_3 \mid \psi \rangle^* 
e^{-2i\chi_0}. 
\end{equation}
In this representation, the unphysical complex conjugation compensates 
the opposite signs of the phase shifts generated by the unitary transform 
$\exp(-i \phi \hat{J}_3)$ at the $m_3$ and $-m_3$ components, so that
a phase shift of $\phi$ multiplies both sides of eq.(\ref{eq:m3cond}) with
the same phase factor of $\exp(-i \phi m_3)$.  

To summarize the main result of the above analysis, a state that achieves its
quantum Cramer-Rao bound in a conventional Mach-Zehnder interferometer must
be path-symmetric as defined by eq.(\ref{eq:m3cond}). Since this equation
is invariant under phase shifts, a path-symmetric state achieves the quantum
Cramer-Rao bound at any bias phase, indicating that the experimentally observed
phase sensitivity of path-symmetric pure state should not depend on phase. 
Any phase dependences of sensitivity such as the ones reported in 
\cite{Oka08,Sun08} 
are therefore the result of either non-optimal estimation techniques, or of
experimental deviations from the intended pure state due to decoherence.

%%--discussion of path-symmetry for commonly used states
Interestingly, the states most commonly considered for quantum metrology are already path-symmetric. Path-symmetry is a natural property of
phase sensitive states since there is no reason to prefer one path over 
another. One important class of path-symmetric states used for metrology is
the class generated by mixing two independent single mode states, 
$\mid \sigma_1 \rangle$ and $\mid \sigma_2 \rangle$, at the input
ports of the beam splitter. In that case, the amplitudes 
$\langle m_1 \mid \psi \rangle$ of the $N$-photon components are equal to
products of the amplitudes of the photon number components of the two
states. According to condition (\ref{eq:m1cond}), path symmetry is 
automatically obtained when both states can be written in terms of real 
amplitudes $\langle n \mid \sigma_{1/2} \rangle = 
\langle n \mid \sigma_{1/2} \rangle^*$. This condition can be fulfilled
by using coherent states, squeezed states, or photon number states. 
Thus the states generated by putting coherent light 
into one port of the interferometer and squeezed vacuum in the other
are all path-symmetric, as are the pair states generated by interfering two 
squeezed vacuum states. Additionally, new types of phase sensitive 
path-symmetric states 
could be generated by putting photon number states into one port and
coherent states or squeezed vacuum states in the other. 

%%--comment on Pezze and Smerzi

We can thus conclude that the quantum states used for two-path interferometry
are usually path-symmetric and therefore achieve their Cramer-Rao bound
in conventional photon counting experiments at any phase. 
The significance of this result is that it greatly simplifies the analysis of
phase sensitivities for two-path interferometry with non-classical states.
In particular, it shows that the recently derived result \cite{Pez08} that
the phase sensitivity of squeezed-coherent light is independent of bias phase
and that it achieves its Cramer-Rao bound applies to all other 
path-symmetric states as well. It is therefore sufficient to determine the
phase sensitivity of conventional photon detection experiments with 
path-symmetric input states directly from the quantum Cramer-Rao bound of
\begin{equation}
\label{eq:CR}
\frac{1}{\delta \phi^2} = 4 \langle \psi \mid \hat{J}_3^2 \mid \psi \rangle.
\end{equation}
Although the previous discussion assumed a fixed photon number $N$, its 
application to fluctuating photon numbers is straightforward, since photon
counting measurements can distinguish subspaces with different photon numbers
$N$. Thus eq.(\ref{eq:CR}) can be applied directly to the phase sensitivity
of path-symmetric states with fluctuating photon numbers such as the ones considered in \cite{Pez08}. The only difficulty that arises from dealing with
fluctuating photon numbers is that the proper identification of the maximal 
phase sensitivity cannot be obtained from the usual Heisenberg limit
of $\delta \phi=1/N$, since $N$ should be the precise number of photons used
in a single phase estimate. Pezze et al. tried to solve this problem by 
combining several measurements into one, reducing the total photon number
fluctuations at the expense of additional shot noise caused by the independent measurements \cite{Pez08}. However, eq.(\ref{eq:CR}) suggests a more 
direct definition 
of the ultimate quantum limit of phase sensitivity for fluctuating photon
numbers. For an $N$-photon state, the Heisenberg limit is obtained from the
maximal value of the path uncertainty $\langle \hat{J}_3^2 \rangle=N^2/4$ 
achieved by a maximally path entangled (NOON) state. If photon numbers fluctuate, the maximal value of $\langle \hat{J}_3^2 \rangle$ is obtained by averaging
over the {\it squared} photon numbers,
\begin{equation}
\label{eq:newHL}
\frac{1}{\delta \phi^2} \leq \langle \hat{N}^2 \rangle.
\end{equation} 
Thus the correct form of the Heisenberg limit is actually higher than the 
square of the average photon number, indicating that the maximal phase 
sensitivity will be underestimated if photon number fluctuations are neglected.

The closeness of a non-classical state to the Heisenberg limit can be
determined from the ratio of $4 \langle \hat{J}_3^2 \rangle$ and
$\langle \hat{N}^2 \rangle$. For the combination of coherent light and
squeezed vacuum discussed in \cite{Pez08}, the result can be obtained
from the coherent amplitude $\alpha$ and the squeezing factor of $\exp[2r]$.
In the limit of high photon number, the result depends only on the 
ratio of photon number averages in the two ports, 
$q=4 \alpha^2/\exp[2r]$. Interestingly, the maximum
of $4 \langle \hat{J}_3^2 \rangle/\langle \hat{N}^2 \rangle$ is obtained
at $q=\sqrt{3}$, and not at equal intensities ($q=1$) as assumed 
in \cite{Pez08}. The phase sensitivity at $q=\sqrt{3}$ is 
$2/(\sqrt{3}+1)\approx 0.73$ times the Heisenberg limit of 
$\langle \hat{N}^2 \rangle$, indicating that the majority of the $N$-photon
components generated come close to achieving the phase sensitivity of
maximally path entangled states \cite{Pez08,Hof07}. 
This result can be compared with the phase sensitivity of the
$N$-photon pair states generated by interfering two squeezed vacuum states.
For a squeezing factor of $\exp[2 r] \gg 1$, the phase sensitivity is
$4 \langle \hat{J}_3^2 \rangle = \exp[4r]/4$ and the Heisenberg limit is
given by $\langle \hat{N}^2 \rangle=\exp[4r]/2$. Thus, the phase sensitivity
is $0.5$ times the Heisenberg limit, lower than the maximum of $0.73$ achieved 
by having coherent light in one input port of the interferometer.

%%--note on single photon state 
It may also be interesting to compare the strategy of increasing phase
sensitivity by squeezing the vacuum in the ``empty'' port of the interferometer
with the alternative of exchanging it with a well defined photon number
state. For a high amplitude coherent state in the other port, the increase
in phase sensitivity given by eq.(\ref{eq:CR}) only depends on the increase of
quantum fluctuations in the quadrature component that determines the 
interference term $\hat{J}_3$ between the coherent amplitude and the modified
vacuum. Hence, using a photon number state is the equivalent of a squeezing
factor of $\exp[2 r]=2n+1$. For a single photon, this is equivalent to the
effect of 4.8 dB squeezing.

%%--note on experiment
From the experimental side, the consequence of the above result is that
the complete output statistics of phase sensitive states should be measured
and evaluated. In previous approaches like the ones reported in
\cite{Oka08,Sun08}, only a specific photon number distribution was measured in
the output, resulting in a theoretical phase dependence of sensitivity 
for the ideal pure state. If the complete output data is used, the phase
dependence of sensitivity originates only from experimental imperfections
and provides an insight into the robustness of the quantum state against
decoherence and noise. Interestingly, an estimate of the robustness may be
obtained from the (phase dependent) estimator values $\lambda A_m$ defined by 
eq.(\ref{eq:ev}). Specifically, the estimator amplifies any background
noise in the measurement result $m$ by a factor of $\lambda^2 A_m^2$. 
Therefore, a convenient definition of the phase dependent robustness $R$ of a 
path-symmetric pure state can be obtained by taking the inverse 
of the sum of the squared estimator values, 
$R=1/(\sum \lambda^2 A_m^2)$. This robustness is a phase dependent feature 
that characterizes and distinguishes different path-symmetric states.
It may thus be a useful tool for the practical optimization of quantum 
metrology strategies.

%%%--Conclusions
In conclusion, the result that conventional photon counting based two-path
interferometry achieves the quantum Cramer-Rao bound for all path-symmetric
pure states allows a generalization of the specific results in 
\cite{Pez07,Pez08} to include the majority of states considered for 
quantum metrology. In essence, this means that it is enough to determine
the path uncertainty $4 \langle \hat{J}_3^2 \rangle$ of a path-symmetric
pure state to obtain the phase sensitivity obtained with that state
at any bias phase. Quantitative predictions for experiments are greatly
simplified, and states approaching the ultimate limit of phase estimation 
given by the generalized Heisenberg limit of eq.(\ref{eq:newHL}) can be
identified and evaluated more efficiently. Thus the concept of 
path-symmetry may pave the way for further progress in the field
of quantum metrology.

%%\section*{Acknowledgment}
Part of this work has been supported by the Grant-in-Aid program of the
Japanese Society for the Promotion of Science.

\end{document}